Title: **Paradoxical LTP maintenance with inhibition of protein synthesis and the proteasome suggests a novel protein synthesis requirement for early LTP reversal**

Authors: Paul Smolen, Douglas A. Baxter, and John H. Byrne

23 pages, 5 figures

Laboratory of Origin:

Department of Neurobiology and Anatomy

W. M. Keck Center for the Neurobiology of Learning and Memory

McGovern Medical School of the University of Texas Health Science Center at Houston

Houston, Texas 77030

Running Title: Paradoxical LTP maintenance suggests novel protein synthesis requirement

Correspondence Address:

Paul Smolen

Department of Neurobiology and Anatomy

McGovern Medical School of the University of Texas Health Science Center at Houston

Houston, TX 77030

E-mail: Paul.D.Smolen@uth.tmc.edu

Voice: (713) 500-5601

FAX: (713) 500-0623

## ABSTRACT

The transition from early long-term potentiation (E-LTP) to late long-term potentiation (L-LTP) is a multistep process that involves both protein synthesis and degradation. The ways in which these two opposing processes interact to establish L-LTP are not well understood, however. For example, L-LTP is attenuated by inhibiting either protein synthesis or proteasome-dependent degradation prior to and during a tetanic stimulus (e.g., Huang et al., 1996; Karpova et al., 2006), but paradoxically, L-LTP is not attenuated when synthesis and degradation are inhibited simultaneously (Fonseca et al., 2006). These paradoxical results suggest that counter-acting 'positive' and 'negative' proteins regulate L-LTP. To investigate the basis of this paradox, we developed a model of LTP at the Schaffer collateral to CA1 pyramidal cell synapse. The model consists of nine ordinary differential equations that describe the levels of both positive- and negative-regulator proteins (PP and NP, respectively) and the transitions among five discrete synaptic states, including a basal state (BAS), three states corresponding to E-LTP (EP1, EP2, and ED), and a L-LTP state (LP). An LTP-inducing stimulus: **1)** initiates the transition from BAS to EP1 and from EP1 to EP2; **2)** initiates the synthesis of PP and NP; and finally; **3)** activates the ubiquitin-proteasome system (UPS), which in turn, mediates transitions of EP1 and EP2 to ED and the degradation of NP. The conversion of E-LTP to L-LTP is mediated by the PP-dependent transition from ED to LP, whereas NP mediates reversal of EP2 to BAS. We found that the inclusion of the five discrete synaptic states was necessary to simulate key empirical observations: **1)** normal L-LTP, **2)** block of L-LTP by either proteasome inhibitor or protein synthesis inhibitor alone, and **3)** preservation of L-LTP when both inhibitors are applied together. Although our model is abstract, elements of the model can be correlated with specific molecular processes. Moreover, the model correctly captures the dynamics of protein synthesis- and degradation-dependent phases of LTP, and it makes testable predictions, such as a unique synaptic state (ED) that precedes the transition from E-LTP to L-LTP, and a well-defined time window for the action of the UPS (i.e., during the transitions from EP1 and EP2 to ED). Tests of these predictions will provide new insights into the processes and dynamics of long-term synaptic plasticity.

**INTRODUCTION**

Blocking the ubiquitin-proteasome system (UPS) blocks long-term spatial memory consolidation in mice (Artinian et al., 2008) and hippocampal-dependent long-term inhibitory avoidance memory in rats (Lopez-Salon et al. 2001) (for review see Fioravante and Byrne, 2011). Proteasome inhibition also inhibits late LTP (L-LTP) of the Schaffer collateral CA3-CA1 pathway (Dong et al. 2008; Fonseca et al. 2006; Karpova et al., 2006). Here, L-LTP is defined as a long-lasting increase in synaptic weight (EPSP amplitude or slope) that lasts for at least 3 h post-stimulus. Similar definitions of L-LTP have been used by others (*e.g.*, Lynch, 2004). In addition, inhibition of protein synthesis (PSI) inhibits L-LTP (Abraham and Williams, 2008; Frey et al., 1998; Stanton and Sarvey, 1984). Together, these data indicate that L-LTP requires both synthesis of proteins, which likely stabilize synaptic potentiation, and degradation of proteins, which in some manner inhibit potentiation.

However, Fonseca et al. (2006) found that L-LTP was preserved when the protein synthesis inhibitor anisomycin was applied concurrently with the proteasome inhibitor lactacystin, with both applications beginning prior to a tetanic stimulus (see also Dong et al. 2008 who reached similar conclusions). These studies used intact hippocampal slices, but in isolated dendrites, L-LTP is also maintained in the presence of proteasome inhibition and anisomycin (Dong et al. 2014). A conceptually similar phenomenon occurs in an associative learning task in *Aplysia* (Lyons et al. (2016). Proteasome inhibition blocked formation of long-term memory (LTM) with no effect on intermediate-term memory (ITM). Paradoxically, concurrent inhibition of the proteasome and PSI did not block LTM. These observations suggest that the two inhibitors applied concurrently appear to cancel each other's effect in some manner. Under these conditions, L-LTP can, paradoxically, be maintained despite apparent inhibition of protein synthesis.

To investigate the bases and implications of these data, we developed a relatively abstract model based on discrete synaptic states, such as non-potentiated and potentiated. Similar state-based synaptic models have proven useful previously. For example, a state-based model has simulated and predicted dynamics of synaptic tagging and capture of plasticity-related proteins (Barrett et al. 2009). In our model, a synapse can be in either a basal state, a state corresponding to early LTP (E-LTP) that does not require protein synthesis, a subsequent state in which proteasome activity alters the E-LTP state to a distinct state that can incorporate protein necessary for consolidation of L-LTP, and a state corresponding to L-LTP. The positive protein is necessary for the transition to L-LTP. The negative protein accelerates the reverse transition from E-LTP back to the basal state. If synthesis of this protein is blocked, E-LTP is greatly prolonged. In addition, degradation of the negative protein is more sensitive to proteasome inhibition than is degradation of the positive protein. Although the model is abstract, we provide a discussion of the potential molecular underpinnings of components of the model (see Discussion).

With these assumptions, we were able to simulate the preservation of L-LTP when PSI and lactacystin were applied concurrently. This preservation of L-LTP was found to be sensitive to the time of proteasome inhibition. If instead of concurrently, proteasome inhibition began 30 min after PSI, L-LTP was strongly inhibited.

## METHODS

Simulations used the forward Euler method. For the simulations of Fig. 2, results with the fourth-order Runge-Kutta method (Burden and Faires, 2005) were compared to verify there was no significant difference in the results. The time step was 40 ms. Prior to stimuli, model variables were equilibrated for at least one simulated day. Programs are available upon request. In Supplementary Material online, a Java program is given that generates the time courses in Figs. 3A1-A3. The model, with standard parameter values that simulate tetanic LTP, has also been deposited in the online database ModelDB.

### List of Abbreviations

| | |
|------|------|
| BAS | Basal state of synaptic strength. |
| EP1 | Early-LTP synaptic state 1. The initial potentiated synaptic state. |
| EP2 | Early-LTP state 2. A longer lasting early state that requires synthesis of a stabilizing molecular factor. |
| STAB | The molecular factor (unspecified, but could be an activated kinase) that stabilizes EP2. |
| ED | Early-LTP state 3. A synaptic state for which some protein degradation (D), mediated by the ubiquitin-proteosome system, has taken place. This degradation is essential to allow the synapse to incorporate new protein and progress to late LTP. |
| LP | Late-LTP synaptic state. |
| NP | Negative protein, that acts to reverse EP2 back to the basal synaptic state BAS |
| PP | Positive protein, that incorporates into or interacts with the ED state to allow it to transform to the LP state |

UPS  Ubiquitin-proteosome system (pathway for protein degradation)

PSI  protein synthesis inhibitor

LAC  lactacystin, an irreversible inhibitor of the ubiquitin-proteasome system

LYS  lysosomal pathway for protein degradation

E-LTP  Early LTP

L-LTP  Late LTP

ITM  Intermediate-term memory

LTM  Long-term memory

## RESULTS

### Development of a model of late LTP induction that combines potentiating and de-potentiating proteins with discrete synaptic states

The model, illustrated in Fig. 1, is relatively abstract in that the molecular identities of specific proteins and synaptic states are not specified in detail. However, candidate molecules for mediating these synaptic state transitions are discussed below (see Discussion). Late LTP (L-LTP) is induced by a strong stimulus (STIM) with duration 20 min, during which STIM has a unitless amplitude of 1. Outside this interval STIM = 0. The model does not describe activation of biochemical signaling pathways essential for L-LTP, such as MAP kinase activation, therefore the 20 min duration of STIM represents implicitly the activation of these pathways. STIM has many sites of action, which are indicated by green lines/outlines in Fig. 1. STIM drives the synaptic state from basal (BAS) to an early potentiated state (EP1) corresponding to E-LTP. EP1 can revert to basal, but a strong stimulus is assumed to increase the level of a stabilizing factor (STAB) that drives the synapse from EP1 to a second potentiated state (EP2).

In contrast, rapid decay of protein-synthesis independent early LTP induced by weak stimuli is assured by postulating that the transition from EP1 to EP2 is driven by the factor STAB, with STAB not being synthesized in response to weak stimuli, but only in response to strong stimuli. With this model architecture, weak stimuli can only drive the synaptic state from basal to EP1. Irrespective of protein synthesis or of its inhibition, EP1 will then decay back to the basal state over a time of ~2 h, characteristic of the decay of early LTP.

EP1 and EP2 are subject to UPS activity, which is modeled as a transition to a synaptic state ED (denoting early potentiation with UPS-mediated degradation). UPS activity is increased by stimuli. STIM also increases synthesis of a positive protein (PP) that is needed for L-LTP, modeled as a facilitator of the transition from ED to a synaptic state LP (denoting L-LTP). EP2 can revert to the basal state. However, this decay is very slow unless stimulus-induced synthesis of a negative protein NP takes place. Thus, the model incorporates a novel protein synthesis requirement for early LTP reversal. Degradation of NP is primarily mediated by the UPS with a small constitutive component. Degradation of PP, in contrast, is entirely constitutive (e.g., through the lysosomal pathway). The state ED can also reverse to basal, on a time scale characteristic of decay of E-LTP. Finally, L-LTP decays very slowly.

Although transcription is necessary under some conditions to sustain L-LTP (Nguyen et al., 1994), the model developed here, and its application to the experiments of Fonseca et al. (2006), can be conceptualized as applying to dendritic processes only, including protein synthesis in the vicinity of active synapses, without input from the nucleus. This restriction to dendrites is plausible because L-LTP in isolated pyramidal cell dendrites can persist for at least 5 h (Vickers et al., 2005), and in isolated dendrites, proteasome inhibition blocks specifically the late phase of LTP (Dong et al., 2008). Protein concentrations are unitless because the identities of PP and NP are not specified.

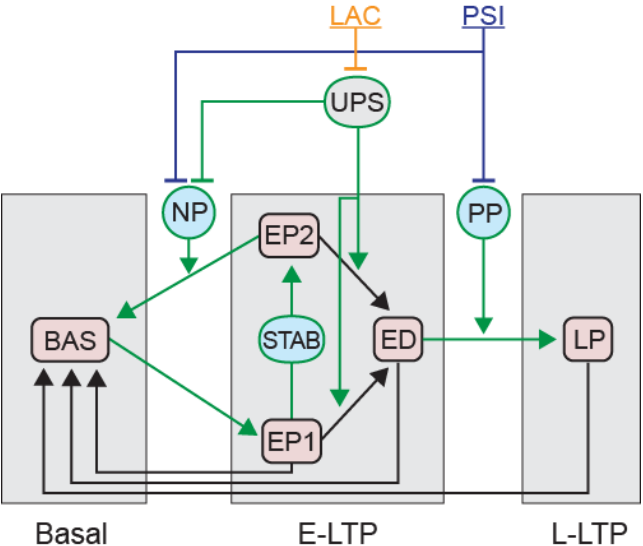

FIGURE 1. Schematic of a model to simulate paradoxical, protein synthesis-independent L-LTP. A LTP-inducing stimulus (STIM, not shown) drives a transition of a synapse from a basal state, BAS, to a labile early potentiated state (EP1). The sites of action for STIM are indicated by green lines/outlines. EP1 either decays back to BAS or transitions into a more stable potentiated state (EP2). Transitions from EP1 to EP2 depend on STIM-induced activation of a stabilizing factor (STAB). Both EP1 and EP2 are subject to degradation by the ubiquitin-proteasome system (UPS), which drives transition to a synaptic state ED. STIM also enhances synthesis of a 'positive' protein PP, necessary for normal L-LTP which requires protein synthesis. PP drives a transition from ED to synaptic state LP. EP2 can decay to BAS. However, this decay rate is very slow unless stimulus-induced synthesis of a de-potentiating or negative protein (NP) occurs. Degradation of NP is mostly via the UPS. Degradation of PP, in contrast, is entirely constitutive. ED can also revert to BAS, on a time scale characteristic of decay of E-LTP. L-LTP decays very slowly (with a very small rate constant) via transition from LP to BAS. The actions of inhibitors are illustrated. Lactacystin (LAC) inhibits the ability of the UPS to both degrade NP and allow the transitions to ED from EP1 and EP2, and protein synthesis inhibition (PSI) inhibits the synthesis of NP and PP.

The parameter LAC denotes the degree of inhibition of the proteasome by lactacystin, PSI denotes the degree of protein synthesis inhibition. Fonseca et al. (2006), applied inhibitors (LAC or PSI) beginning 40 min prior to the LTP induction. Empirically, the percentage of inhibition is commonly ~95% or higher (Milekic and Alberini 2002; Squire and Davis 1975). Thus in our simulations, PSI equals 0 until 40 min prior to stimulus, then PSI increases to 0.95. PSI remains high for the remainder of the simulation, because Villers et al. (2012) reported that anisomycin continues to inhibit protein synthesis by ~90% at 4 h after application. Lactacystin is an irreversible inhibitor that covalently modifies the proteasome (Fenteany et al. 1995). Thus in simulations, LAC equals 0 until 40 min prior to stimulus onset, then immediately and irreversibly increases to 0.95. STIM activates UPS, which then decays back to basal activity. The ordinary differential equations (ODEs) for the proteins PP and NP, the stabilizing factor STAB, and UPS activity are as follows,

$$\frac{d[PP]}{dt} = (1 - PSI)(k_{syn1} STIM + k_{syn1bas}) - k_{deg1}[PP] \qquad 1)$$

$$\frac{d[NP]}{dt} = (1 - PSI)(k_{syn2} STIM + k_{syn2bas}) - k_{deg2} UPS[NP](1 - LAC) - k_{deg2bas}[NP] \qquad 2)$$

$$\frac{d(STAB)}{dt} = k_{syn3} STIM + k_{syn3bas} - k_{deg3} STAB \qquad 3)$$

$$\frac{d(UPS)}{dt} = k_{act} STIM + k_{actbas} - k_{deact} UPS \qquad 4)$$

The sum of the synaptic state variables is conserved at 1, corresponding conceptually to a deterministic model of changes in a population of individual synaptic contacts affected by a given stimulus. The state variables track the fractions of the population that are in particular states. Thus, the basal synaptic state BAS = 1 − EP1 − EP2 − ED − LP. The ODEs for the synaptic states are as follows,

$$\frac{d(EP1)}{dt} = k_{f1} STIM(BAS) + k_{f1bas}(1 - STIM)(BAS) - k_{f2}(UPS)(1 - LAC)(EP1)$$
$$- k_{b1}(EP1) - k_{f3}(STAB)(EP1) \qquad 5)$$

$$\frac{d(EP2)}{dt} = k_{f3}(STAB)(EP1) - k_{f4}(UPS)(1 - LAC)(EP2) - k_{b2}(EP2)[NP] \qquad 6)$$

$$\frac{d(ED)}{dt} = k_{f2}(UPS)(1\text{-}LAC)(EP1) + k_{f4}(UPS)(1-LAC)(EP2)$$
$$- k_{b3}(ED) - k_{f5}[PP]^2(ED)$$

(7)

$$\frac{d(LP)}{dt} = k_{f5}(ED)[PP]^2 - k_{b4}(LP)$$

(8)

$$\frac{d(BAS)}{dt} = -k_{f1}STIM(BAS) - k_{f1bas}(1-STIM)(BAS) + k_{b1}(EP1)$$
$$+ k_{b2}(EP2)[NP] + k_{b3}(ED) + k_{b4}(LP)$$

(9)

In Eqs. 7 and 8, the transition rate from the state ED to the state LP uses the second power of the positive protein concentration PP. Because the state LP corresponds to normal L-LTP, this second power nonlinearity represents, in a simple and abstract way, the necessary cooperative interaction of multiple proteins to remodel and strengthen a synapse that undergoes L-LTP.

The synaptic weight W is expressed as a linear combination of the synaptic states, with potentiated synaptic states having higher individual weight coefficients (EP1 and EP2, corresponding to E-LTP, and LP, corresponding to L-LTP) as follows,

$$W = BAS + 5(EP1) + 8(EP2) + 4(ED) + 8(LP)$$

(10)

Because the individual synaptic states are not empirically characterized, the coefficients in Eq. 10 are simply chosen so that the time courses and magnitudes of simulated L-LTP in the four conditions of no inhibitor, PSI only, LAC only, and PSI and LAC concurrently, are similar to data. With no inhibitor, strong stimuli lead to a long-lasting increase in LP specifically. From Eq. 10, a long-lasting increase in LP corresponds to a long-lasting increase in W above its basal level, and therefore to observed L-LTP.

Standard values for the rate constants in the model are given in the following table, along with the reaction or synaptic state transition the parameter is associated with. Because concentrations are unitless, the only parameter unit is time, taken as minutes.

| Parameter name | Value | Associated Reaction |
|---|---|---|
| $k_{syn1}$ | 1.0 min$^{-1}$ | Synthesis of PP |
| $k_{syn1bas}$ | 0.0035 min$^{-1}$ | Synthesis of PP |
| $k_{deg1}$ | 0.005 min$^{-1}$ | Degradation of PP |
| $k_{syn2}$ | 2.0 min$^{-1}$ | Synthesis of NP |

| $k_{syn2bas}$ | 0.0014 min$^{-1}$ | Synthesis of NP |
|---|---|---|
| $k_{deg2}$ | 0.01 min$^{-1}$ | Degradation of NP |
| $k_{deg2bas}$ | 0.002 min$^{-1}$ | Degradation of NP |
| $k_{syn3}$ | 1.0 min$^{-1}$ | Synthesis of STAB |
| $k_{syn3bas}$ | 0.008 min$^{-1}$ | Synthesis of STAB |
| $k_{deg3}$ | 0.02 min$^{-1}$ | Degradation of STAB |
| $k_{act}$ | 0.2 min$^{-1}$ | Activation of UPS |
| $k_{actbas}$ | 0.00214 min$^{-1}$ | Activation of UPS |
| $k_{deact}$ | 0.0143 min$^{-1}$ | Deactivation of UPS |
| $k_{f1}$ | 0.1 min$^{-1}$ | BAS → EP1 |
| $k_{f1bas}$ | 0.001 min$^{-1}$ | BAS → EP1 |
| $k_{f2}$ | 0.02 min$^{-1}$ | EP1 → ED |
| $k_{b1}$ | 0.005 min$^{-1}$ | EP1 → BAS |
| $k_{f3}$ | 0.01 min$^{-1}$ | EP1 → ED |
| $k_{f4}$ | 0.02 min$^{-1}$ | EP2 → ED |
| $k_{b2}$ | 0.0007 min$^{-1}$ | EP2 → BAS |
| $k_{b3}$ | 0.02 min$^{-1}$ | ED → BAS |
| $k_{f5}$ | 0.0005 min$^{-1}$ | ED → LP |
| $k_{b4}$ | 0.001 min$^{-1}$ | LP → BAS |

For this abstract model, parameter values were not chosen to fit biochemical data, because the molecular processes mediating synaptic state transitions are not specified. The standard parameter values were chosen by trial and error to allow concurrent simulation of the qualitative dynamics of four cases: normal LTP, LTP inhibition by LAC and by PSI individually, and LTP restoration in the presence of LAC and PSI as in Fonseca et al. (2006). There is no suggestion that this value set is unique in its ability to do so.

**The model simulates normal LTP and attenuation of LTP by individual inhibitors**

Figure 2 illustrates L-LTP in the absence of inhibitors. Prior to stimulus, 3000 min were simulated to ensure a stable baseline for all variables (i.e., time = −3000 to 0 of which only 50 min are illustrated). At time = 0, the stimulus was applied and an additional 300 min were simulated. The E-LTP states, EP1 and EP2, increase rapidly and then decrease over 1-2 h (Fig. 2A). STAB, NP, and PP also increase (Fig. 2B). The increase in ED, due to proteasome-mediated degradation, is transient, because PP drives ED to LP (Fig. 2C). The synaptic weight W increases rapidly upon stimulus, corresponding to E-LTP (states EP1 and EP2), and then remains elevated, corresponding to L-LTP (state LP).

In Fig. 2C, as in all simulations of L-LTP, after the synaptic strength W reaches its maximum, a very slow decay of W over h is evident. L-LTP was not modeled as a permanent switch to a stable upper state of W. We incorporated this slow decay to represent the consistent empirical slow decay of normal L-LTP that is evident in two of the studies that illustrate (at least partial) preservation of L-LTP with concurrent proteasome inhibition and PSI (Dong et al. 2008, 2014). In the third study, little if any decay is evident over ~2.5 h (Fonseca et al. 2006). Nonetheless, for an abstract model that qualitatively simulates the dynamics of synaptic strength in these three studies, keeping a slow decay process in the model appears the most parsimonious choice.

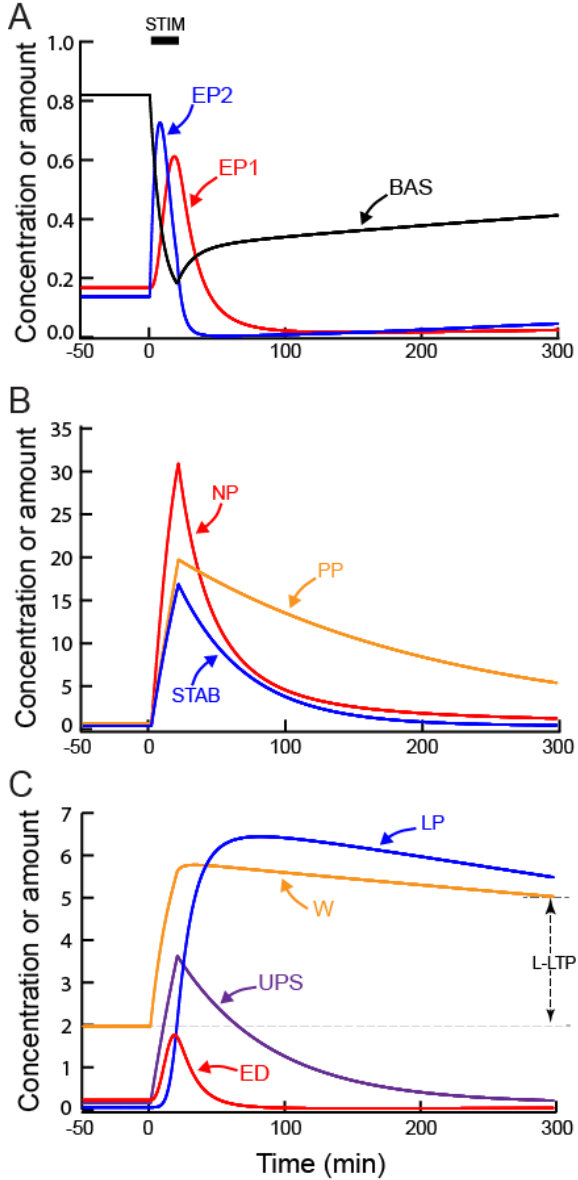

FIGURE 2. Response of the model to an LTP-inducing stimulus. (**A**) STIM drives increases in EP1 and EP2, which then decay over ~1-2 h. The decay of EP2 requires the NP protein. The actions of STIM result in a long-lasting decrease in the basal synaptic state BAS. (**B**) STIM also drives increases in the levels of the stabilizing factor STAB and the proteins NP and PP. (**C**) The increase in the synaptic state ED, following UPS activation, is transient, because the PP protein drives a transition of the ED state into the long-lasting LP state. To allow visualization of the dynamics, some time courses were vertically scaled as follows: in (**A**) the EP1 and EP2 time courses were scaled by 2, in (**C**) the ED and LP time courses were scaled by 10. As indicated by the dashed line, the magnitude of L-LTP is defined as the increase in W at the end of the simulation.

Figures 3A1-A3 illustrate the attenuation of LTP when <u>only</u> the proteasome is inhibited by lactacystin (LAC). E-LTP is robust as seen by large increases in EP1 and EP2. LAC, by inhibiting the proteasome, slows the degradation of NP (compare with Fig. 2), but not PP. Due to LAC, the ED state hardly increases because the UPS-mediated transitions from EP1 and EP2 are inhibited, and only a small subsequent increase in the L-LTP state, LP, occurs. Therefore, for the synaptic weight W, only early, not late, LTP occurs. Inhibition of UPS increases the level of NP, which in turn reverses EP2 to the basal synaptic state BAS, terminating E-LTP.

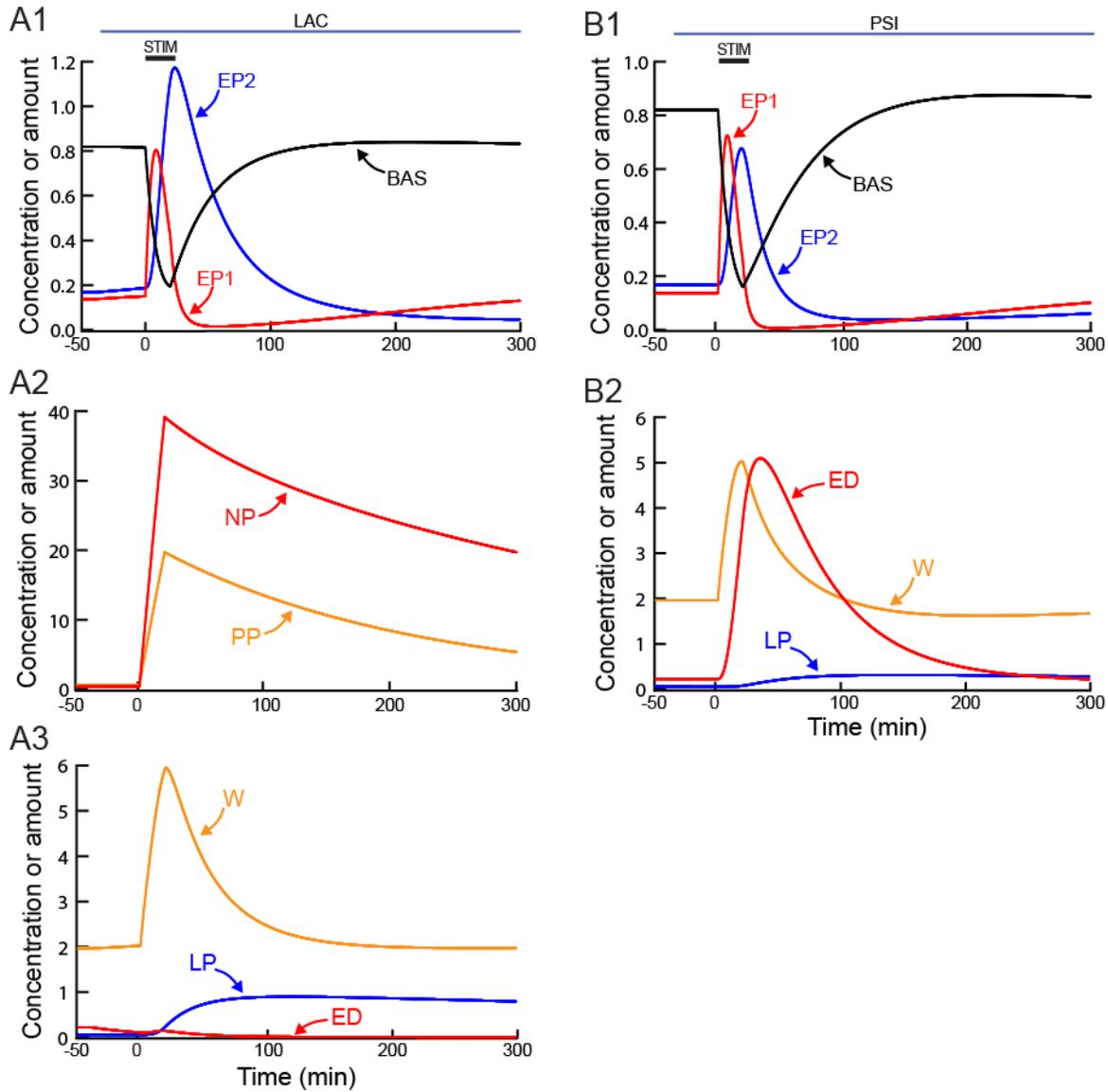

FIGURE 3. Stimulus responses with inhibitors. **(A1 – A3)** Response when degradation by the UPS is inhibited. Inhibition of the UPS (indicated by the bar labeled LAC) was begun 40 min prior to STIM and continued throughout the simulation. **(A1)** E-LTP is robust as seen by large increases in EP1 and, especially, EP2. However, with UPS inhibition, the decrease in BAS is no longer persistent. **(A2)** Compared to the control simulation (Fig. 2B), UPS inhibition slows the degradation of NP, but does not alter PP degradation. **(A3)** Due to UPS inhibition, STIM induces only a small increase in the level of the synaptic state ED. With ED small, only a small increase in the LP state, corresponding to protein synthesis-dependent L-LTP, occurs. This small increase is balanced by the decay of EP1 and EP2 so that, for the weight W, no late potentiation is evident. Only E-LTP is seen. **(B1 – B2)** Response when protein synthesis is inhibited (indicated by the bar labeled PSI). **(B1)** EP1 and EP2 increases are still robust. **(B2)** UPS still generates the state ED. However, because synthesis of PP is inhibited, the increase in the state LP is greatly reduced. For the synaptic weight W, E-LTP is preserved but L-LTP is reduced. EP1, EP2, ED, and LP were vertically scaled as in Fig. 2.

Figures 3B1-B2 illustrate the attenuation of L-LTP when protein synthesis is inhibited. EP1 and EP2 increases (E-LTP) are robust. Because PP synthesis is inhibited, the increase in LP is substantially reduced compared to the control (Fig. 2). The strong PSI applied here (95%) causes a small decrease in basal synaptic weight in this model at later times. For $t > 2$ h after the start of PSI, W decreases below its value prior to PSI.

**Simulated restoration of L-LTP by an inhibitor combination**

The model simulates paradoxical L-LTP, or maintenance of elevated synaptic weight W, given concurrent UPS inhibition and PSI. In the model, this synaptic weight increase corresponds not to protein-synthesis dependent L-LTP, but to a pronounced prolongation of E-LTP, by prolonging the lifetime and therefore the elevated level of the specific synaptic state EP2 (Fig. 4A). This prolonged elevation is due to two factors, 1) EP2 cannot readily reverse to the basal state because synthesis of the negative protein NP is inhibited, and 2) EP2 also cannot progress to ED because UPS is inhibited. PSI also strongly attenuates stimulus-induced increases in PP and in NP, comparing the peak values in Fig. 4B with those in Fig. 2B. This attenuation prevents progress to LP and also slows reversal to the basal state BAS.

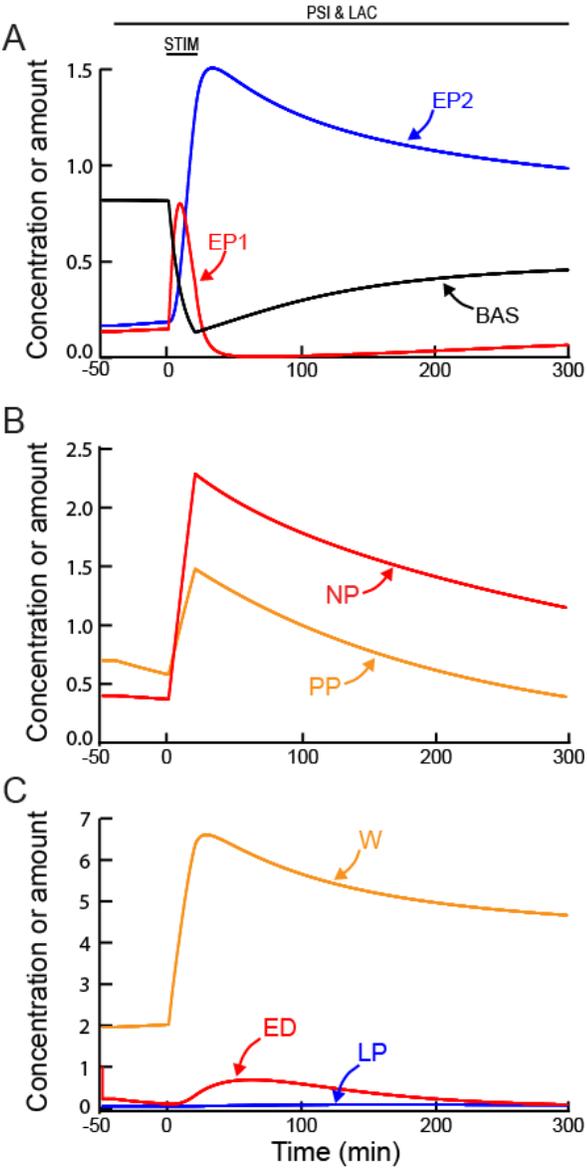

FIGURE 4. Simulation of the paradoxical induction of L-LTP with concurrent PSI and LAC. (**A**) The bar labeled PSI & LAC indicated the time course of inhibition of protein synthesis and degradation. The increase in EP2, corresponding to E-LTP, is greatly prolonged. (**B**) Stimulus-induced increases in PP and NP. PSI attenuates these increases (compare with Fig. 2). The decreased level of NP is responsible for the prolonged lifetime of EP2. (**C**) Due to UPS inhibition and PSI, levels of ED and LP are greatly reduced. Nonetheless, the synaptic weight W exhibits L-LTP, due to the pronounced prolongation of the lifetime of EP2. ED, LP, EP1, and EP2 are vertically scaled as in Fig. 2.

Figure 4C illustrates that increases in the LP and ED states are greatly reduced (compared to Fig. 2). Nonetheless, the synaptic weight W exhibits long-lasting LTP. This is due to the great prolongation of the lifetime of EP2.

LAC is an irreversible proteasome inhibitor. Therefore we conducted a further simulation to determine if paradoxical L-LTP would decay if proteasome inhibition was reversed subsequent to stimulus. The simulation of Fig. 4 with PSI and LAC applied concurrently was repeated except that LAC was set to zero at 100 min post-stimulus. In response, the synaptic weight subsequently decayed to the basal level, albeit slowly (over ~200 min).

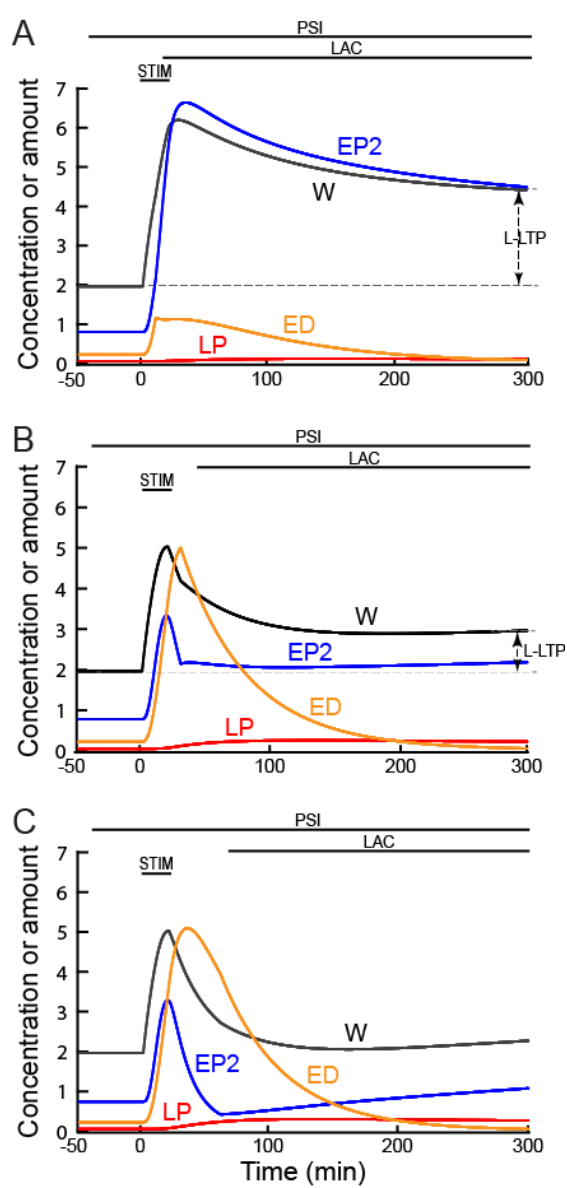

## Blocking L-LTP with delayed onset of proteasome inhibition

A model prediction is that the preservation of L-LTP by dual PSI and LAC would be progressively lost when the time of onset of LAC is delayed to follow STIM. In Figs. 5A-C, LAC onset is delayed by respectively 10, 30, and 60 min after the onset of STIM. For 10 min (**A**), there is not enough time post-stimulus for the proteasome to convert the synaptic state EP2 to ED, prior to LAC onset. ED and LP remain at low levels, and EP2 is persistent as in Fig. 4, resulting in apparent L-LTP (W time course).

FIGURE 5. Time window for the effects of UPS inhibition. To examine the time at which LAC was most effective, the simulated application of LAC was delayed after the onset of PSI and STIM. PSI begins 40 min prior to stimulus (indicated by bar labeled PSI). UPS inhibition (as indicated by bars labeled LAC) is delayed relative to STIM onset by (**A**) 10 min, (**B**) 30 min, and (**C**) 60 min. (**A**) With a 10 min delay, dynamics are similar to Fig. 4. ED and LP remain at low levels. The time course of EP2 is greatly prolonged, resulting in apparent L-LTP (time course of W). Relative to Fig. 4, some decrease in L-LTP is evident even with the 10 min delay. (**B**) With a 30 min delay, prolonged elevation of EP2 is still evident, but is strongly attenuated, because substantial UPS activation now occurs, transforming EP2 to ED. Correspondingly, W still exhibits apparent L-LTP, but substantially reduced. (**C**) With a 60 min delay, EP2 decays to baseline. The late UPS inhibition causes a subsequent small increase. The increase in LP is minor because PSI inhibits PP

synthesis.  Thus for W, L-LTP is strongly inhibited compared to normal (Fig. 2), and is similar to PSI alone (Fig. 3).  Vertical scaling as in Fig. 2 except EP2 is scaled by 10.

For 30 min post-stimulus (**B**), the persistent elevation of EP2, and of W, is substantially smaller because there is now time for the proteasome to transform EP2 into ED, which can revert to BAS.  For 60 min, (**C**) the UPS is active long enough to drive EP2, *via* transfer to ED, all the way to the pre-stimulus baseline.  However, although ED is generated, the increase in LP is minor, because PSI inhibits PP synthesis.  L-LTP is substantially reduced.

## DISCUSSION

A novel form of L-LTP, stable in the presence of concurrent inhibition of the proteasome and of protein synthesis, is modeled as an expression of stabilized E-LTP.  The absence of E-LTP reversal in the presence of PSI is modeled as a novel requirement for protein synthesis (the protein NP) for E-LTP reversal.  The model makes the empirical prediction that some synaptic changes that are likely to require protein synthesis, such as expansion of the PSD (Bosch et al. 2014), will not occur with this stabilized E-LTP. The model, and the data simulated, follow L-LTP induced by tetanic 100 Hz stimuli for up to 180 – 200 min post-tetanus (Dong et al. 2008; Fonseca et al. 2006).  At yet later times (~4 h) the maintenance of L-LTP is also blocked by transcription inhibitors (Frey et al. 1996; Vickers et al. 2005), and it is likely, although not modeled here, that maintenance of stabilized E-LTP would similarly depend on transcription at these times (note: other authors reported that tetanic L-LTP was affected earlier by transcription inhibitors, after ~ 2 h, Kelleher et al. 2004; Nguyen and Kandel 1994).  In addition, Dong et al. (2008, their Fig. 8B) provide evidence that L-LTP induced by theta-burst stimuli may be more sensitive to transcription inhibition than is tetanic L-LTP, with a partial decay at ~3 h.  Therefore it would be of interest to examine whether theta-burst stimuli, concurrent with both PSI and proteasome inhibition, would result in stabilized E-LTP as robust and long-lasting as that reported by Fonseca et al. (2006) and simulated here.

In addition, the model predicts that this stabilized E-LTP will destabilize, and revert to basal synaptic strength, when proteasome inhibition is removed, because then the E-LTP synaptic state (EP2 in the model) can convert to the state after proteasome action (ED), which in turn decays more readily to the basal state. Fonseca et al. (2006) used the irreversible proteasome inhibitor LAC.  Using a reversible proteasome inhibitor such as MG-132, concurrently with PSI, could therefore provide a key test of the model.  Long-lasting LTP should still be induced, but if the activity of MG-132 reverses over h, then decay of this LTP is predicted to occur as proteasome activity resumes.

Ligation of ubiquitin to proteins destined for degradation is an essential process within the proteasome pathway. Numerous E3 ubiquitin ligases, with differing substrate specificities, are responsible for this process. To further delineate some of the molecular mechanisms by which proteasome activation supports L-LTP, it may prove useful to use small interfering RNA to knock down expression of specific E3 ubiquitin ligases suggested to be important for synaptic function or for learning and memory, and then to characterize the synaptic substrates of those ligases found to be necessary for L-LTP. For example, HERC 1 ligase is already known to be important for associative learning (Perez-Villegas et al. 2018), Ube3A is known to be important for learning and L-LTP (Jiang et al. 1998), and Mib1 is important for dendritic spine outgrowth (Mertz et al. 2015).

An early hypothesis (Dong et al., 2008) suggested that proteasome inhibition impairs L-LTP maintenance by blocking degradation of the transcription factor ATF4, which is a repressor of transcription mediated by cAMP response element binding protein (CREB). If this occurred, ATF4 levels would build up, preventing transcription of CREB-activated genes necessary for L-LTP maintenance. This hypothesis could explain the dynamics of L-LTP that is dependent on transcription at relatively early times and is therefore not rescued by simultaneous application of proteasome inhibitor and PSI. In particular, theta-burst L-LTP is not fully rescued by this combined application (Fig. 8B of Dong et al. 2008) and may therefore depend on ATF4 degradation in this way. In contrast, for L-LTP induced by tetanic stimuli, this hypothesis is not compatible with the observed preservation of L-LTP by pairing proteasome inhibition with PSI (Fonseca et al. 2006), because in that case PSI would repress translation of the mRNAs resulting from CREB-activated transcription, and then concurrent proteasome inhibition would only tend to increase whatever residual level of ATF4 remained, thus decreasing transcription of these mRNAs, and further reducing levels of products of CREB-activated transcription.

More recently Dong et al. (2014) suggested the following alternative model based on processes occurring in the vicinity of activated synapses. LTP induction is accompanied by, and late LTP requires, relatively rapid synthesis or activation of translational activators, which then support synthesis of proteins required for L-LTP maintenance. Concurrently, slower basal synthesis of proteins that function as translational repressors is ongoing. If proteasome activity is blocked, the levels of translational repressors build up, until they become able to suppress ongoing translation necessary for L-LTP maintenance, and L-LTP therefore is not maintained. An implicit assumption here is that the levels or efficacy of translational activators do not build up to the same extent as do the repressors with proteasome inhibition, perhaps because the activators are degraded predominantly through a non-UPS pathway.

The above constitutes a plausible alternative hypothesis for the efficacy of proteasome inhibition in blocking late LTP. However, this conceptual model as it stands has difficulty explaining the preservation of L-LTP maintenance in the presence of concurrent proteasome inhibition and protein synthesis inhibition. With concurrent inhibition, it is still plausible that translational activators could be phosphorylated or otherwise activated in response to stimulus – but with protein synthesis inhibited, this should not suffice, because these activators would not be able to induce synthesis of proteins required for late LTP maintenance. The dynamics of the proteasome and of translational repressors would then appear not likely to be relevant. It is possible that in a mathematical model based on these assumptions, kinetic parameters could be fine-tuned (e.g., with partial inhibition of proteasomes and of protein synthesis, and with nonlinear, or saturable, effects of proteins on L-LTP maintenance) to simulate some extent of late LTP with both inhibitors present. However, it does not appear likely that such a kinetic model would provide a robust qualitative explanation for the maintenance of all or most of the initial amplitude of LTP, with little decay observed, in the corresponding experiments of Fonseca et al. (2006). In the data of Dong et al. (2014), the proportion of peak LTP that was maintained after ~3 h with both inhibitors present appears to be substantially reduced relative to normal L-LTP, and this extent of preservation might be possible to simulate with such a kinetic model.

A later intriguing result indicates that synaptic protein degradation by the UPS may depend on ongoing NMDA receptor (NMDAR) activity (Fonseca, 2012). In this study, anisomycin was applied during tetanus, and the NMDAR antagonist AP-5 was applied immediately after tetanus. Late LTP, instead of decaying due to PSI, was preserved by the presence of AP-5. A hypothesis to explain this preservation, in the context of our model, is that AP-5 leads to inhibition of protein degradation by the UPS, possibly by blocking a component of $Ca^{2+}$ influx through the NMDAR that is necessary for UPS activation. In corroboration of this hypothesis, AP-5 was found to inhibit degradation of a reporter protein in neurons stimulated with bicuculline (Djakovic et al. 2009). If this hypothesis is correct, the synapse would be unable to transfer to the intermediate state ED, because in our model, transition to ED requires UPS activation (Fig. 1). Thus, the synapse would remain in the potentiated state EP2. With protein synthesis also inhibited, the negative protein NP is not synthesized. Therefore, the state EP2 decays only very slowly back to the basal state. Thus, the synapse remains in the potentiated state EP2 for a long time, corresponding to apparent preservation of L-LTP in the presence of concurrent PSI and AP-5. In this case, it is plausible that, as discussed above for concurrent anisomycin and LAC application, structural synaptic changes characteristic of normal L-LTP would not be observed.

In the model, after consolidation of normal L-LTP in the absence of inhibitors, the synapse is in the state LP, which is assigned a very slow rate constant (long time scale) for decay back to the basal state BAS. If this parameter value choice is correct, then the model predicts that application of LAC alone, or of AP-5 alone, during the maintenance phase of normal L-LTP, with application not beginning until after L-LTP consolidation, would only very slowly reverse L-LTP, over a longer time scale than the experiments of Fonseca et al. (2006, 2012). This prediction follows from the kinetic scheme of Fig. 1 in which, irrespective of the dynamics of protein synthesis or the UPS, the only pathway available to leave the state LP is the slow decay to BAS.

Considering possible molecular correlates of synaptic states in the model, exocytosis of AMPA receptors (AMPARs) from intracellular storage into the postsynaptic density (PSD) is a plausible candidate for the stabilizing factor STAB to generate the state EP2 from EP1, corresponding to E-LTP that takes longer to reverse and can progress to L-LTP. As discussed in Penn et al. (2017), blocking this exocytosis by tetrodotoxin allows short-term potentiation only, abolishing LTP. Blocking only surface diffusion of AMPARs, but allowing exocytosis, allows for development of a reduced component of E-LTP, demonstrating that the rate of exocytosis is in fact increased after stimulus. In this case the species STAB (Eq. 6) would not represent the level of a molecule, but the activity of AMPAR exocytosis. An alternative candidate stabilizing factor is the activity of the guanine nucleotide exchange factors Kalirin and/or Trio. These factors are activated by CaM kinase II, enhance postsynaptic AMPAR incorporation, and are both necessary and sufficient to allow LTP induction (Herring and Nicoll, 2016; Xie et al. 2007).

In contrast, a protein that enables endocytosis of AMPARs out of the PSD would be a candidate for the negative protein NP, which in the model enhances reversal of E-LTP. Expression of NP should also be enhanced by LTP-inducing stimulus. A protein that has been commonly studied to elucidate its relationships to synaptic plasticity, the Arc/Arg3.1 protein, appears to fulfill both these requirements. Arc enhances AMPAR endocytosis (Wall and Corrêa, 2017), and high-frequency stimulus strongly induces Arc mRNA expression, which then localizes to dendritic regions containing activated postsynaptic sites (Steward et al., 1998). A complicating factor is that Arc synthesis is also required for late LTP in the CA3-CA1 pathway and in dentate gyrus (Messaoudi et al. 2007; Plath et al. 2006). Thus, in the context of the model, Arc may also function as a positive protein (PP). This putative dual function for Arc is plausible, because recent data suggest that post-translational SUMOylation of Arc can switch Arc between two functional states. Non-SUMOylated Arc may promote activity-dependent AMPA receptor endocytosis, while SUMOylated Arc appears to promote late LTP via regulation of the actin cytoskeleton (Carmichael and Henley 2018; Craig et al. 2012; Nair et al. 2017).

In the model, there is a cascade of four synaptic states that is traversed to establish usual L-LTP in the absence of inhibitors (BAS → EP1/2 → ED → LP). With protein synthesis present, as a synapse progresses from EP1/2 to LP, the states become more difficult to reverse to BAS. This cascade resembles the potentiated states in the model suggested by Fusi et al. (2005), which would constitute a synaptic architecture favorable for increased memory storage capacity, while preserving ease of memory induction. Further investigation of this similarity may be of interest.

Finally, Lyons et al. (2016) found in *Aplysia* that rapamycin inhibited LTM but that rapamycin in combination with proteasome inhibition did not inhibit LTM. This paradoxical effect has similarities to the restoration of L-LTP by pairing anisomycin with LAC, because rapamycin also inhibits protein synthesis, by inhibiting mTOR-activated translation. However, in *Aplysia*, rapamycin also inhibits persistent MAP kinase (MAPK) activation (Michel et al. 2011), and MAPK activity is known to be necessary for synaptic long-term facilitation (Martin et al. 1997; Sharma and Carew 2004). In mammalian cells, rapamycin has been reported to induce expression of mitogen-activated protein kinase phosphatase 1 (Rastogi et al. 2013), which could inhibit any late phase of MAPK activation. Activity of the ERK isoforms of MAPK is known to be necessary for L-LTP (English and Sweatt 1997; Rosenblum et al. 2002). Taken together, these observations indicate it would be of interest to repeat the Fonseca et al. (2006) experiment, replacing anisomycin with rapamycin. A finding that proteasome inhibition was also able to rescue L-LTP inhibition when applied concurrently with rapamycin, would suggest that the resulting L-LTP, which our model suggests is due to stabilization of early LTP by absence of the negative protein NP, is independent of not only protein synthesis but also persistent ERK MAPK activation. However, to confirm this, the effect of rapamycin on late ERK activation in this system would need to be assessed.

**SUPPLEMENTARY MATERIAL (2 files, online)**

Supplementary.doc – MS Word file describing the supplementary Java program.

Fig3A13.java – Java program for simulating the time courses shown in Figs. 3A1-A3.

**ACKNOWLEDGEMENTS**

We thank Ashok Hegde and Harel Shouval for comments on an earlier version of the manuscript. Supported by NIH grant NS102490.